\documentclass[preprint,review,12pt]{elsarticle}
\usepackage{graphicx}
\usepackage{amssymb}
\usepackage{natbib}
\biboptions{square,comma,numbers,sort}

\def\({\left(}
\def\){\right)}
\def\[{\left[}
\def\]{\right]}
\def\<{\left<}
\def\>{\right>}
\def\ave#1{\<{#1}\>}

\journal{Physica D, to appear in 2021 as paper 133012}

\begin{document}

\begin{frontmatter}

% Include your paper's title here
\title{Dynamics of contentment} 

\author{A.A. Burluka}

\address{Faculty of Engineering \& Environment, Northumbria University,\\
  Newcastle-upon-Tyne, NE1 8ST, The United Kingdom }

\date{Received: / Accepted }

\begin{abstract}
This work formulates a mathematical model aimed at prediction of temporal evolution of joint probability density of wealth and contentment in a society. A continuous variable changing between 0 and 1 is introduced to characterise  contentment, or satisfaction with life, of an individual and an equation governing its evolution is postulated from analysis of several factors likely to affect the contentment. As contentment is strongly affected by material well-being, a similar equation is formulated for wealth of an individual. Subsequently, an evolution equation for the joint probability density of individuals' wealth and contentment within a society is derived from these two equations and an integral representation of marriage effects. As an illustration of this model capabilities, effects of the wealth tax rate over a long period of time are simulated for a society with an initially low variation of wealth and contentment: the model predicts that a higher taxation in the longer run may lead to a wealthier and more content society. It is also shown that lower rates of the wealth tax lead to pronounced stratification of the society in terms of both wealth and contentment and that there is no direct relationship between the average values of the latter two variables thus providing an explanation to the Easterlin paradox.

\end{abstract}


\end{frontmatter}

%\linenumbers

\section*{Introduction}

Taken at a level of an individual, the satisfaction with life, i.e. oneself, own environment, current economical situation, and many other factors some of which are fleeting and minute, may be considered as a strongly fluctuating quantity. There exist several measures of individual well-being and life satisfaction formulated in distinctly different ways. The simplest measure is when an individual grades his own happiness, satisfaction with this or that factor or life in general, on a prescribed scale; the resulting value is usually coarsely-grained and  it is usually difficult to discern how it is influenced by individual subjective or objective factors. This difficulty prompted development of more complicated measures such as the TSWLS ( Temporal Satisfaction With Life Scale) \citep{TSWLS}, a sum of grades given to several questions about life satisfaction in the past, present and future. The different questions have an equal weight and the outcome is an integer variable taking values between 15 and 105 and a number of works demonstrated that TSWLS averages are comparable to other similar questionnaire-based measures. Another example of life satisfaction measures is the National Time Accounting, or the U-index, where U stands for ``{\it unpleasantness or the fraction of time spent in an unpleasant state or activities}'': the monograph \citep{Uind} provides a very comprehensive introduction of this measure, its connection to other measures of the well-being, and a critical appraisal of assumptions on which it is based. It should be noticed that the calculation of the U-index does introduce weights of different activities taken as some subjective ``intensity of unpleasantness'' thus facilitating quantitative analysis of how different factors affect the life satisfaction. There are also  numerous other criteria of subjective well-being often lacking precise mathematical definition, see e.g. \citep{Mod3P} and references therein.

It should be noticed that all these criteria are, by construction, average quantities with an averaging period which may vary between several hours and many years. It is hardly surprising that attempts to establish a statistical correlation of such parameters with objective factors affecting an individual met only partial success, \citep{Kahn06}.  Typically, a model aimed at predicting such a ``happiness'' factor would be a more or less sophisticated algebraic regression, \citep{Kops19}, combining subjective assessment of one own well-being with external factors affecting the latter. At the same time, neglect of the subjective factors is commonly made in the economic theories and it usually leads to a somewhat reductionist view of the individual life satisfaction assuming that it is determined by the utility functions to which an individual would necessarily seek an extremum, see e.g. \citep{Fala09,Pugno13} as representative samples of this approach. In this economics-based framework, life satisfaction goes hand in hand with  the level of consumption and increasing financial wealth, yet there is ample evidence \citep{East10} that this is not so and there are some strong transient effects in that relationship: this essentially constitutes the well-known Easterlin's paradox. Very lucid and concise critical comparison of the economics and psychology approaches to description of the well-being may be found in \citep{East03}.

Yet, it is obvious that, regardless of how it may be quantified, an individual's happiness is not an entirely subjective variable as it is also strongly influenced by external agents, e.g. the current state of economy or level of cultural activity. What is more,  presence of intrinsic fluctuations in one's personal life satisfaction strongly suggests the need for an investigation at the level of a ``society'', that is an ensemble of individuals sharing the same daily activities, an ensemble sufficiently large so as to admit a statistical description. Ultimately, it is the ensemble statistical properties of a suitably quantified life satisfaction which matter for a society: as an example, the important societal issue of growth of inequality cannot be described in terms of individual or even average well-being as it requires at least knowledge of the second-order moments, some sort of ``root-mean-square happiness''. Thus, there is a clear need for a quantitative description of distribution of life satisfaction in a society and response of this distribution to changes in individuals' circumstances: this works aims exactly at developing such a description.

Rather than building on one of the existing definitions for individual happiness departing from subjective factors, it seems convenient to introduce a new variable and rather than trying to define it in terms of subjective perceptions of an individual, this new variable is defined through an evolution  equation taking into account the major factors affecting  well-being. The term ``contentment'' for this new variable is used to distinguish it from the already existing measures of well-being such as U-index or TSWLS, or various utility functions use of which is firmly associated with variational problems \citep{Sher20}. Essentially, the contentment is defined here as a solution of an evolution equation postulated for an individual as an ordinary differential equation (ode) including a number of factors.

Rather intensive literature search revealed only one attempt \citep{Sprott04}
to formulate a mathematical model for temporal evolution of individual's happiness. That work \citep{Sprott04} took  as a departure point a simple linear dynamic system proposed in \citep{Stro94} for evolution of affection in a couple, the description later developed in \citep{Rinaldi2013_1,Rinaldi2013,Rinaldi2014}.
Arguing that the happiness should be proportional to the accumulation of true feelings, \citep{Sprott04} postulated a couple of non-linear oscillator model equations for it and while their solutions exhibit a variety of regimes, their realism is difficult to gauge owing to lack of connection between those model equations and any observable quantity. In contrast with that approach, every term in the evolution equation developed in this work for contentment represents an observable and eventually measurable factor, the characteristic time of which may be identified and estimated.

Life satisfaction depends upon, among other factors, and to a large extent, the objective quality of life and its subjective perception by an individual. Both objective life quality and its perception depend on the relative standing of this individual within the society and the rate at which this standing improves; one may assume that the latter is correlated with the self-fulfilment. To characterise this standing a variable representing wealth is introduced; while it is most readily associated with monetary wealth and such an association is used here for mathematical expression of specific factors affecting the dynamics of contentment, the notion of wealth may be generalised to  include other factors affecting the individual's standing in the society, e.g. level of skills or some other non-pecuniary forms of societal recognition. 
Similarly to the contentment, the temporal evolution of the wealth of an individual is described in terms of an ordinary differential equation. The ode's for the wealth and contentment of an individual are coupled and from them is derived a partial differential equation (pde) for the joint probability density function (pdf) of the contentment and wealth within the society. However, following this approach, the effects of marriage, essentially a pairwise interaction, cannot be easily included in the ode's for an individual's wealth and contentment and these effects are represented as an integral term entering directly the joint pdf equation.

Owing to the lack of the fundamental conservation laws for the wealth and contentment, the mathematical model for these variables may only be postulated but not derived; the approach adopted here is to approximate the
rate of a particular effect with a simplest mathematical expression compatible with everyday experience. The reasoning here is done in terms of characteristic times of each factor affecting wealth or contentment. Values of adjustable parameters entering such an expression are chosen so that, taken individually, the effect in hand produces the expected rate of change of wealth or contentment. This approach is entirely empirical and ad-hoc, however, it allows, firstly, a simple comparison of the magnitude of individual effects, and secondly, a straightforward means of improving every expression when additional data becomes available.  

\section*{Development of the model}

Let $C$, a real variable varying from 0 to 1, denote the contentment so that a perfectly unhappy person be attributed a zero value and someone at the state of perfect bliss unity.  Clearly, a unit of contentment may be established by attributing a specific value of $C$, say between 0 and 1, taken as degree of satisfaction with life resulting from a particular objective and observable event, however, this is not pursued here. Use of continuous rather than discrete variable is convenient as it allows subsequent comparisons with diverse previous measures ranging from 3-point scale of \citep{East03} to 105-point TSWLS  \citep {TSWLS}. The unit of time, $t$, is taken here as one year.  

While the real variable quantifying ``wealth'' $M$ is not, {\it sensu stricto}, bounded from either above or below, as one may, at least in theory, have an infinitely large debt or fortune, it is taken here as non-negative, i.e. debt is neglected. Its unit is taken here as half of the average household wealth as it makes it easier to formulate the model so that it yields the temporal evolution of wealth compatible with the relatively well documented trends, e.g data from the Bank of England, \citep{BoE1}, or Office for National Statistics in the UK, \citep{ONS1}. For the numerical simulations illustrating the model, it is taken bounded from above by a positive value, $M_{max}$, of the average wealth of the upper decile in the UK in 2018. 

\subsection*{Evolution of contentment of an individual}
It is assumed that the rate of change of contentment for every individual comes from several additive factors, namely, satisfaction with own wealth $W_1$ and income $W_2$, satisfaction with the infrastructure of the society $W_3$, quality of the neighbourhood $W_4$, and dissatisfaction with high taxes $W_5$ and inequality in the society $W_6$. Besides, there are two additional terms, one, $W_7$ describing the fact that everyone is seeking and finding satisfaction in things independent of the wealth and another $W_8$ representing rapidly changing and statistically uncorrelated events affecting contentment. The rate of evolution of the individual contentment is then simply a sum of these factors:
\begin{equation}
\frac{dC}{dt} = \sum_i W_i 
\label{dCdt}
\end{equation}
Notice that the above ode does not include change of contentment resulting from marriage, or formation of family, which is treated later. These factors cover a mix of subjective, objective, direct, and indirect aspects \citep{WB1}.  

Contentment due to wealth increases or decreases when the wealth is greater or smaller than the average and, in absence of other factors, it usually takes about several years before this difference of wealth is considered as permanent and it leads to a lasting improvement of own satisfaction with life. Therefore, one may use a simple linear expression for $W_1$:
\begin{equation}
W_1( M ) = K_1 ( M - \ave{M} ) 
\label{W1}
\end{equation}
where $K_1$ is a reciprocal of the characteristic time for this factor and the angular brackets denote average for the society. As postulated, it is obvious that, should both the individual wealth $M$ and the society average wealth $\ave{M}$ stay constant, this term alone would lead to a change of the individual contentment proportional to time as $C(t)=C(t=0) + K_1 ( M - \ave{M} )t$. For an approximation of $K_1$, one may think that material comfort of life for someone with twice the average wealth would lead to a perfect contentment within 10 years, thus $K_1\approx 0.1$ -- this value is retained in what follows.

That the contentment should increase with income may sound controversial, e.g. it has been argued that income has no influence as may be derived from data obtained when tracing satisfaction with life of differently educated groups of population, taking the education for proxy measure of income, see e.g. Fig. 4 in \citep{East03}. However, the same data do show a significant increase in contentment for both the more and the less educated groups at ages where one may expect a rapid career growth with stagnation afterwards corresponding again to an age where most careers tend to achieve a plateau. Besides, it seems natural that not all sources of individual's income contribute equally to contentment: it is difficult to imagine that the income from welfare contributes to satisfaction with life even though it may cover the material necessities. While the analysis of influence of income on happiness show that it changes with the age of an individual \citep{East03}, the age distribution of the society is not considered in the present model. Furthermore, what matters is not the magnitude of income as such but the excess of the disposable income $I_d$ over the cost of living $L$, or what is generally perceived to be a ``good income'' $G$. Thus, one may tentatively suggest:
\begin{equation}
W_2( M, C ) = K_2 ( I_d(M, C) - G ) 
\label{W2}
\end{equation}
A specific expression for the disposable income $I_d(M,C)$ is formulated later; the reciprocal of the characteristic time for this effect $K_2\approx 0.1-0.2$, i.e. the estimation that a very comfortable income takes 5 to 10 years to bring a full swing of contentment; in what follows the value $K_2=0.15$ is used.

The term related to the satisfaction with the infrastructure $W_3$ at this level of generalisation should reflect the fact that it does not depend on the individual circumstances but on the general level of the expenditure of the society on the common use facilities such as education, culture and sport institutions. Rather importantly, it is this term here which includes investment of the society into the health care, thus indirectly it also includes the important contribution of good health into the contentment. All such expenses are assumed to come from the general taxation and their benefits spread evenly among the individuals; this choice is an idealisation reflecting the UK NHS model but more refined expressions may be easily formulated for applications of the model to other circumstances. Thus
\begin{equation}
W_3 = K_3 \alpha_I T_t 
\label{W3}
\end{equation}
where $T_t$ is the total tax intake in the society and $\alpha_I$ is the fraction of it going into infrastructure. Reflecting the fact that the infrastructure projects have relatively long time scales of order of 10-20 years, the estimation for the product $\alpha_I K_3\approx 0.05$ may be made.

As the taxation is always positive, the above term leads to a uniform increase of $C$ while the dissatisfaction with the high taxation $W_5$ should decrease $C$ in proportion to the tax paid by the individual $T(M,\dot M(M,C))$ which may depend on the individual's wealth $M$ and total gross income $\dot M(M,C)$:
\begin{equation}
W_5 = - K_5 T( M, \dot M(M,C) ) 
\label{W5}
\end{equation}
The characteristic time of this term should be of order of the time between the general elections, at least for the democratic societies as the pledge to decrease taxes often proves to be an election winner: this fact indicates that the dissatisfaction with the tax level may overtake other considerations\footnote{One of the reviewers raised the point that the terms describing contentment with the infrastructure and discontent with the taxation should be related as one may expect that taxes cause less discontent if they are seen as well spent. It is correct up to a point, but the fact nonetheless remains that the tax burden on the individuals is concentrated at the upper deciles of the wealth distribution, at least in the OECD countries, while the benefits of the infrastructure are spread more or less uniformly across the society. Therefore, keeping these terms separate would enable the model to simulate the differential rates of contentment change produced across the different strata of the population.} 
  As the overall tax burden in the OECD countries is somewhere between 30-60\%, the $K_5$ value should be somewhere in the range 0.3 to 0.7; $K_5=0.5$ may thus be suggested.

The ``quality'' of the neighbourhood is an integral characteristic of the society, and the average contentment $\ave{C}$ is taken here as its proxy, making straightforward to approximate the term accounting for change of satisfaction from communal relationships:
\begin{equation}
W_4 = K_4 \( \ave{C} -\frac{1}{2} \) 
\label{W4}
\end{equation}
This expression also subsumes the stronger tendency of the more contented society to take better care of its members as compared with a society in strife characterised by low values of $\ave{C}$.

Along with the well-known Gini coefficient, the wealth inequality, i.e. the width of the marginal probability density $P(M)$ in the society may be characterised by the root-mean-square (rms) value \citep{Klya05}:
$$\ave{M'^2}^{1/2}=\( \int_0^{M_{max}} (M-\ave{M})^2 P(M)dM \)^{1/2} $$
where $M_{max}$ is the upper bound, possibly infinite, on the wealth in the society; prime is used herebeneath to denote the deviation from the mean value. However, it is not the rms magnitude of the wealth of society per se but rather its ratio to the mean wealth, or ``intensity of the inequality'' which seems to cause the discontent. While there seems to be no study directly supporting this assumption, it may be inferred from the fact that, while the inequality in the OECD countries is at the same level as in the 1930's \citep{PikeCap}, now owing to increase in average wealth this ratio is lower and the large-scale economic protest movements and trade unions are much weaker than they were back then; there is certainly no economic movement in any OECD country now even remotely approaching Jarrow Crusade in the UK or French Front Populaire in scale and mobilisation of the society. Thus one may suggest for the term $W_6$ describing effects of discontent caused by inequality:
\begin{equation}
W_6 = - K_6 \frac{\ave{M'^2}^{1/2}}{ \ave{M} } 
\label{W6}
\end{equation}
where $K_6\approx 0.05 \div 0.1$ as the time scale for this effect is 10$\div$20 years as evidenced by the timescales for raise and fall of large-scale social protest movements. $K_6=0.1$ is retained here.

Regardless of wealth or income, everyone seeks solace finding it in various things, so that it may be assumed that there is a general tendency of increase of contentment and this increase is the larger the smaller is the contentment, decreasing to zero at $C=1$. In common terms, one would seek and often find an inner peace and the term $W_7$ corresponding to this would include effects of religion, hobbies, and voluntary work for good causes,and other feel-good factors independent of material circumstances. The simplest form of $W_7$ reflecting this would be:
\begin{equation}
W_7 = K_7 \( 1 - C \)^\beta 
\label{W7}
\end{equation}
where $\beta > 1$ to ensure the smoothness of $W_7$ at $C=1$. The characteristic time scale for this term depends on the probability distribution of $C$ but if to consider that it takes several, say between 2 and 10, years for someone initially at full discontent to find solace, $K_7 \approx 0.1 \div 0.5$; the values of $K_7=0.15$ and $\beta=2$ are adopted in what follows.

Finally, there are always fluctuations in anyone's satisfaction of life the time scale of which is very short in comparison with any term introduced above. These may be caused by multitude of events, big and small, ranging from a burnt toast to a life-changing lottery win with no causal relationship between them. Some of those events decrease and some increase contentment but their net effect on the average contentment is zero; they produce rapid random changes of individual's contentment ultimately reducing variation of contentment within the society. Assuming their independence of the individual's circumstances, the contribution of such events to the contentment may therefore be described as a stochastic function of time $\zeta(t)$: 
\begin{equation}
W_8 = \zeta(t)
\label{W8}
\end{equation}
such that $\ave{\zeta(t)}=0$ at any time and
$\ave{\zeta(t)\zeta(t')}=\gamma \delta(t-t')$
where $\delta(t)$ is Dirac's delta-function and $\gamma$ is the magnitude of the fluctuations of the contentment induced by these random events. Owing to the normalisation of $C$ to unity, the characteristic time scale for reduction of the contentment variations is $\gamma^{-1/2}$ \citep{Klya05}. The levelling of the contentment variations caused by this random events happens at the time scale of the same order as the average life expectancy, say 75 years, thus $\gamma\approx 0.11$. One may conclude with the caveat that, obviously, the numerical values of the rate factors in different terms in Eq.~\ref{dCdt} are approximate; moreover, one and same factor would affect different individuals at different magnitude. It is possible to formulate more refined mathematical formulation, for example assuming some statistical distribution of parameter values, but the resulting complexity seems unwarranted at this stage.  

\subsection*{Evolution of wealth of an individual}

Clearly, both the distribution of wealth within the society and the individual's wealth influence the distribution of contentment, thus investigation of the dynamics of the latter should be conducted jointly with the consideration of the temporal evolution of the wealth and its statistical distribution. Even though the ``wealth'' and ``income'' affecting the contentment are not necessarily pecuniary, the model formulation is much easier to perform associating these notions with the aggregated form of monetary wealth making no distinction between various types of financial means, property, pensions etc. The main assumption made here is that the rate of change of an individual's wealth equals the sum of his productivity $U_1$, some amount of the added value $U_2$ produced by others that he alienates from them, taxes $U_3$, and welfare assistance $U_4$. The only coupling between the contentment and the rate of change of wealth may come from the fact that the productivity of an individual with very low contentment is low but is increasing, more or less rapidly, as his contentment rises, whereas, obviously, neither taxes nor welfare payment depend on the contentment.

The dependency of the productivity $U_1(M, C)$ on the contentment $C$ is approximated with a power function reflecting fast drop in productivity as $C$ tends to zero and much slower rate of change as $C$ tends to one:
\begin{equation}
U_1(M, C) = L_1 C^{\alpha_{p1}} M^{\alpha_{p2}} \(M_{max} -M\) ^{\alpha_{p3}}
\label{U1}
\end{equation}
The dependency of the productivity on the wealth $M$ is approximated here with a polynomial such that the productivity has a broad maximum for $M$ in between 1 and 2 and decreases to zero when $M$ tends to either 0 or its maximum value $M_{max}$. For low values of $M$, this functional form reflects the statistical correlation between the destitution and unemployment, ill-health, lack of education and any other factor contributing to low productivity. It should be mentioned that the usual statistical data, see e.g. \citep{ONS1}, clearly demonstrate this correlation between the income and wealth, however, they do not allow one to disentangle easily the individual's own productivity from other contributions to the rate of change of his wealth. The value of the factor $L_1$ governs the overall productivity of the society; in what follows it is fixed such that the productivity of an individual of average wealth $M=1$ is about $0.2M$. The value of $\alpha_{p1}$ determines how steep the decline of productivity is with the discontent: with the value $\alpha_{p1}=1/3$ retained here, with decrease of $C$, decrease of $U_1$  is slow at first, down to approximately 60\% of the maximum when $C=0.2$, followed by a very rapid precipitation to zero for lower $C$. The choice of the power exponents $\alpha_{p2}$ and $\alpha_{p3}$ determine how flat is the dependency of $U_1$ on $M$: for the values adopted here $\alpha_{p2}=1$ and $\alpha_{p3}=3$ the prodictivity stays within 20\% of its maximum value in the interval $0.8\le M\le 2.8$.  

The term $U_2(M)$ describing redistribution of the income within the society towards the elites must necessarily be negative for low $M$ and positive for large $M$, thus there should be a value $M^\star$ such that $U_2(M^\star)=0$ as there is no reason to assume $U_2$ discontinuous. Because this term only redistributes wealth, it must not change $\ave{M}$. Its extremum values may only be attained at the ends of the interval $[0,M_{max}]$. While there is an infinite number of possible functions satisfying these requirements, a simple exponential form is adopted here:
\begin{equation}
U_2(M) = L_2\( \exp\( \frac{M - M^\star}{M_s} \) -1 \)
\label{U2}
\end{equation}
where the parameter $M_s$ determines how fast is the growth of profit from capital. Equation \ref{U2} means that this growth is faster than linear reflecting the correlation between higher levels of investment, risk and return open to the wealthiest \citep{Pike17}. For the developed countries, the ratio between the average income and the nominal gross domestic product per capita is approximately 0.7-0.8 suggesting that $L_2\approx 0.2 \div 0.3$.
For $L_2=0.2$ the value $M_s=2.5$ yields pre-tax rate of the wealth increase of the UK top wealth decile  individuals considered here, $M=7.0$, of approximately 20\% in line with the average values reported for the developed countries, e.g. \citep{Pike17}, but use of Eq.~\ref{U2} with fixed values of $L_2$ and $M_s$ may produce income distribution across society outside the spread observed in the OECD countries in the last sixty years.

Equation~\ref{U2} has three parameters, $L_2, M^\star$ and $M_s$. As this redistribution term should not affect $ \ave{M}$,  $\ave{U_2} = \int_0^{M_{max}} U_2(M) P(M) dM=0$ where
$P(M)=\int_0^1 P(M,C) dC$ is the marginal pdf of the wealth. From this:
$$ M^\star = M_s \ln\ave{\exp\frac{M}{M_s}}
$$
and the values of $L_2$ and $M_s$ are determined so as to satisfy two additional constraints. The first constraint is the prescribed rate of the wealth increase for individuals with $M= M_{max}$; this rate, 12\% per annum is taken as the average rate for the top 1\% derived from the UK statistics, \citep{ONS1}. The second constraint is the alienated added value per worker.  Taking again the UK as the representative of OECD countries and using the Office for National Statistics figures, one see that in 2018 75.8\% of the population were economically active, and the GDP per capita was \pounds 33,140: this means that the productivity per worker was \pounds 43,800. The mean pre-tax income of the worker was \pounds 34,200, thus the fraction of the total GDP redistributed to the wealthiest is approximately 0.22. As $P(M,C)$ evolves, the values of the two constraints $L_2$ and $M_s$ also change and are found iteratively to satisfy these two constraints.

While there exists a great variety of tax regimes, the simplified treatment used here assumes a flat tax on income supplemented with a wealth tax $T_w$ imposed on individuals with the wealth above a threshold $M_w$:
$$ U_3 = L_T ( U_1 + U_2 ) + T_w(M)
$$
where $L_T$ is the total rate of income tax including direct, and indirect, e.g. value-added, taxation. $L_T=0.5$ is taken here as an approximation to the sum of the current basic rate of the UK income tax, National Insurance contribution and Value Added Tax, for a person of average income making little savings. The wealth tax is approximated as:
\begin{equation}
  T_w(M) = \left\{ \begin{array}{lr}  0  &\mbox{ if } M \le M_w \\
   L_3\, \exp\( (M - M_w) (M_{max}-M) \)  &\mbox{ if } M \le M_w  
    \end{array} \right.
\label{TW}
\end{equation}
Even though most OECD countries currently have no direct wealth tax, they do have inheritance taxes and Eq.~\ref{TW} includes those too. Its combination with a flat income tax also describes progressive taxation to some extent, e.g. assuming delayed capital gains tax etc. Equation~\ref{TW} yields progressive decrease of $T_w$ for very large values of $M\approx M_{max}$ representing effects of tax evasion and avoidance by the wealthiest individuals. While it is straightforward to include a progressive income taxation in the description, it is thought that the simplified representation of the taxation split into two parts, one depending on the wealth $M$ and the other on its rate of change, is sufficient for illustration of tax effects on the contentment dynamics. Values of $L_3$ were varied to demonstrate its effects on the temporal evolution of $P(M,C)$.

At the same level of simplification, the welfare payments are assumed here to supplement income of individuals with $M < \ave{M}$ tapering to zero at $M=\ave{M}$:
\begin{equation}
  U_4(M) = \left\{ \begin{array}{lr} L_4\(1 - \frac{M}{ \ave{M} }\)^{\alpha_w}
    &\mbox{ if } M \le \ave{M} \\
    0 &\mbox{ if } M \ge \ave{M} 
    \end{array} \right.
\label{U4}
\end{equation}
where variation of the power exponent $\alpha_w$ achieves representation of the different welfare policies: values of $\alpha_w>>1$ means most of the welfare concentrating around $M\approx 0 $ with steep decrease of $U_4$ for $M\approx \ave{M}$, i.e an alms-type welfare. $1/2 \le \alpha_w \le 1$ produce nearly linear decrease of $U_4$; $\alpha_w =1/2$ is used in what follows. The value of the $L_4$ factor is so determined that $U_4$ compensates exactly all other factors except marriages resulting in zero net change of wealth at $M=0$.

Finally, the rate of change of an individual's wealth may be written as
\begin{equation}
  \frac{dM}{dt}\(M,C\) = \( 1-L_T \) ( U_1( M,C ) + U_2(M) ) - T_w(M) + U_4(M)
\label{Mdot}
\end{equation}
This expression with the values of different parameters specified above produces overall yearly increase of the average household wealth of approximately 8\% in line with the average UK trends over the period 1998-2008.

Following the argument made above that the welfare payments should not directly contribute to increased contentment, the income $I_d$ affecting the contentment as given by Eq.~\ref{W2} may now be written as:
\begin{equation}
 I_d = \( 1-L_T \) ( U_1( M,C ) + U_2(M) ) - T_w(M)
\end{equation}
and the ``good'' income $G$ may simply be taken as the average $\ave{\frac{dM}{dt}}$.

\subsection*{Representation of effects of formation of family/marriage}
From the view of society as an ensemble of individuals and neglecting polygamy, the marriage, or more generally, formation of a family is a pairwise interaction process. In terms of evolution of wealth of individuals this process results in redistribution of wealth of the interacting pair into two equal parts; this leaves the average wealth unchanged and decreases its variance. Effects of marriage on wealth are remarkably similar to the phenomenon of scalar small-scale mixing in a turbulent flow, the subject of numerous investigations, see e.g. \citep{Dopa94}.
Based on this similarly, a convenient starting point for expressing how $P\(M,C,t\)$ is affected by marriages is provided by the so-called integral models of mixing pioneered in \citep{Frost60} and \citep{Curl63}.

There is some evidence that, on average, married individuals show satisfaction with life slightly higher than the celibate ones,  however, this effect seems to be transitional and of the same magnitude as the spread of the ``mean happiness'' see e.g. Fig. 3 in \citep{East03}, therefore, an assumption is made here that the effect of marriage on the contentment is to equipartion it, producing equal probability of any value $C$ regardless of the initial contentment. Secondly, it is assumed that the probability of marriage between two individuals depends only on their wealth but not contentment. The dependency of the probability of marriage on the difference of the wealth of the two individuals is described with a weight function $f$, unity for zero difference, and decreasing to zero when this difference becomes very large. It is thought that introduction of such weight function reflects the general tendency of marriages within the same social strata.
Root-mean-square value of the wealth in the society is chosen as the measure for the disparity of the wealth resulting in decrease of the probability of marriage.
For an individual, the marriage happens with the characteristic time scale $\tau_f$ which is simply the average life expectancy divided by an average number of marriages over the lifetime; the latter is currently about 1 in the UK so $\tau_f\approx 80$.

With these assumptions, the expression for the rate of change of $P(M,C)$, caused by marriages, denoted here as $I_f(M,C)$, may be written as:
\begin{eqnarray}
  I_f(M,C) &=& \displaystyle\frac{P(M,C)}{\int_0^1 dC' P(M,C')} \cdot \frac{1}{\tau_f}  \cdot \nonumber \\
  &  { } & \[ \int_0^M dz f\(z,\ave{M'^2}\) \int_0^1 dC' P(M+z,C') \int_0^1 dC'' P(M-z,C'') \right. \nonumber \\
 &  \qquad & - \left. \frac{ P(M,C) }{4} \int_0^{M_{max}} dz f(z,\ave{M'^2}) \int_0^1 dC' P(z,C') \]
  \label{Ifam}
\end{eqnarray}
where the first integral describes generation of probability at given $M$ and $C$ from all admissible marriages of individuals at any contentment and wealth $M_1$ and $M_2$ such that $M_1+M_2=2M$ and the second term -- removal of states from those marriages. It may be shown that Eq.~\ref{Ifam} preserves normalisation of $P$ and $\ave{M}$ if $\int_0^\infty dz f\(z,\ave{M'^2}\) =1 $. In what follows, exponentially decaying weight function was used:
\begin{equation}
  f\(z,\ave{M'^2}\) =  \( \ave{M'^2} \)^{1/2} \exp\( - \frac{z}{ \( \ave{M'^2}\)^{1/2}} \)
\label{fz}
\end{equation}
and it should be noticed that use of other alternatives for $f$ had not led to qualitatively different results. 

\subsection*{Evolution equation for $P(M,C)$}
The established evolution equations for the individual's contentment, Eq.~\ref{dCdt}, and wealth, Eq.~\ref{Mdot}, allow one to write an equation describing temporal evolution of the joint probability density function (pdf) $P(M,C,t)$ of these two variables for the entire society,  \citep{Moya49,Pop94}, as:
\begin{eqnarray}
  \frac{\partial P \( M,C,t\) }{\partial t} &+& \displaystyle
  \frac{\partial }{\partial M} \ave{\frac{dM}{dt}\(M,C,t\)} P \(M,C,t\)\ 
  \nonumber \\
  &+\ &\frac{\partial }{\partial C} \ave{\frac{dC}{dt}\(M,C,t\)} P \(M,C,t\)\  
  \nonumber \\
  &{ } +& \displaystyle \frac{\partial^2 }{\partial C^2}
  \int_0^t dt'\ave{\frac{dC}{dt}\(M,C,t\) \frac{dC}{dt}\(M,C,t'\) }
  P\(M,C,t\)   \nonumber \\
  \ &=&\   I_f\( M,C,t \)
  \label{P1}
\end{eqnarray}
where the term $I_f\( M,C,t \)$ is given by Eq.~\ref{Ifam}. Equation~\ref{P1} is a truncated Kramers-Moyal equation where the dropped terms are identically zero owing to the structure of Eqs.~\ref{dCdt} and \ref{Mdot}. It should be noticed that formulation of a partial derivatives equation for the probability density is an approach alternative to the population dynamics methods, see e.g. \citep{Landi_fash} in which the pdf carrier is partitioned and an ode for the number of individuals in each partition would be solved: both method would be very similar in computational requirements for one-dimensional pdf, but clearly the latter method would require a very large set of ode to represent a two-dimensional pdf.

The conditional averages in Eq.~\ref{P1} are trivially found  for all terms in Eq.~\ref{Mdot} and all terms in Eq.~\ref{dCdt} but $W_8$. The latter term upon averaging yields only a diffusion-like term with the second partial derivative with respect to $C$. The final equation for  $P(M,C,t)$ thus becomes: 
\begin{eqnarray}
 &{ }& \frac{\partial P \(M,C,t\) }{\partial t}\, + \, \displaystyle
 \frac{\partial }{\partial C} \[ \sum_{i=1}^7 W_i\] P\(M,C,t\) \nonumber \\
 \ &+&\ \displaystyle  \frac{\partial }{\partial M} \[ \( 1-L_T \) ( U_1( M,C ) + U_2(M) ) - T_w(M) + U_4(M)\] P \(M,C,t\)  \nonumber \\
 &\ =&\  \gamma \displaystyle \frac{\partial^2 P\(M,C,t\)}{\partial C^2}\ +\  
  I_f\( M,C,t \)
  \label{P2}
\end{eqnarray}
where $\gamma$ is the rms magnitude of the random events, see Eq.~\ref{W8}.
The boundary conditions imposed on $P$ are zero normal derivatives at $C=0,1$ and $M=0,M_{max}$ lines ensuring the preservation of normalisation of $P$. It should, however, be noticed that non-zero $\frac{dM}{dt}\(M,C,t\)$ and $P$ at $M=M_{max}$ would require dynamic adjustment of $M_{max}(t) ={\rm max}_C \frac{dM}{dt}\(M_{max},C,t\)$ but this is not attempted here. The main reason for keeping large but constant $M_{max}$ is that, as will be seen later, $P\(M,C,t\)$ is small and concentrated in a fairly small vicinity of the point $(M_{max},1)$ affecting little the dynamics of contentment in the regions of large $P$ representing most of the society. The second reason is that numerical methods for computational domains with moving boundaries are more complicated and the extra complexity is not warranted at the stage of the formulation of the model principles. 

\section*{Illustration of the model: evolution of an initially homogeneous society}

\subsection*{Numerical implementation for $P(M,C)$}
The computational domain $C\in[0,1]\, \times M\in[0,M_{max}]$, with $M_{max}=12$,  was covered by uniform grid on which Eq.\ref{P2} was discretised using control volume method with implicit upwind expressions for the fluxes on the boundaries of grid cells, the time derivative was calculated with simple first-order approximation. Averages and root-mean-square values and  the integrals in Eq.~\ref{Ifam} were calculated explicitly using method of overlapping parabolas implemented in DAVINT routine \citep{davint}. The resulting sparse algebraic system of equations was solved using Gaussian elimination with partial pivoting and LU decomposition and preconditioning implemented in SUPERLU-MT set of routines \citep{Superlu}. The sensitivity to the grid size and the time step selected so that equivalent to CFL criterion in fluid mechanics does not exceed 0.8 were verified to ensure that the results were not sensitive to their choice.  

\subsection*{Evolution of $P(M,C)$ for homogeneous society: effects of tax on wealth}
%\subsection*{Initial conditions for $P(M,C)$}

The joint probability density $P(M,C)$ for society completely homogeneous in terms of contentment and wealth is $\delta(M-\ave{M}) \delta(C-\ave{C})$.
Joint action of marriages and various random factors acting on a very short time-scale, rhs of Eq.~\ref{P2}, broadens the initially infinitesimally narrow peak of contentment distribution  into a normal distribution which is then distorted and clipped at the boundaries $C=0,1$ until it attains a constant value of unity between these boundaries. The model postulates that contentment increases the productivity, Eq.~\ref{U1}, thus non-zero $\ave{C'^2}^{1/2}$ will produce broadening of the wealth distribution and non-zero $\ave{M'^2}^{1/2}$: the pdf carrier, i.e. the domain where $P(M,C) \ne 0$, becomes a curve in $M-C$ space. Under wealth-distributing action of the marriages, this $P(M,C)$ carrier curve will spread along constant $C$ lines becoming two-dimensional area. In simple words, this model predicts that  in a society where everyone was initially equally rich or poor and equally satisfied with life, after a certain time there will be a significant probability of meeting an individual at any state of contentment or satisfaction with life together with a certain variation of wealth. It is possible that a particular choice of welfare and tax rates and functional shapes may produce a steady-state $P(M,C)$ but this aspect is beyond the scope of this work. 

In reality, no society is completely homogeneous, thus a ``nearly homogeneous'' society is defined here as having a normal distribution for $P(M,C)$ with non-zero rms values $\ave{M'^2}^{1/2}$ and $\ave{C'^2}^{1/2}$. By construction, the initial value of $\ave{M} =1$, and thus the initial average wealth becomes a unity for scaling. However, the initial value of $\ave{C}$ remains arbitrary because no scale is defined for $C$ between its extreme values of 0 and 1. The initial normal distribution is clipped at the boundaries and renormalised to unity. In what follows, the parameters of the initial $P(M,C)$ are $\ave{M'^2}^{1/2}=0.6$, $\ave{C}=0.5$ $\ave{C'^2}^{1/2}=0.08$; use of different values yields qualitatively similar results. The simulations were run for the time period of 75 units (years), slightly smaller that the current UK life expectancy.

\begin{figure}[ht]
\begin{center}
\includegraphics[width=.95\textwidth,clip]{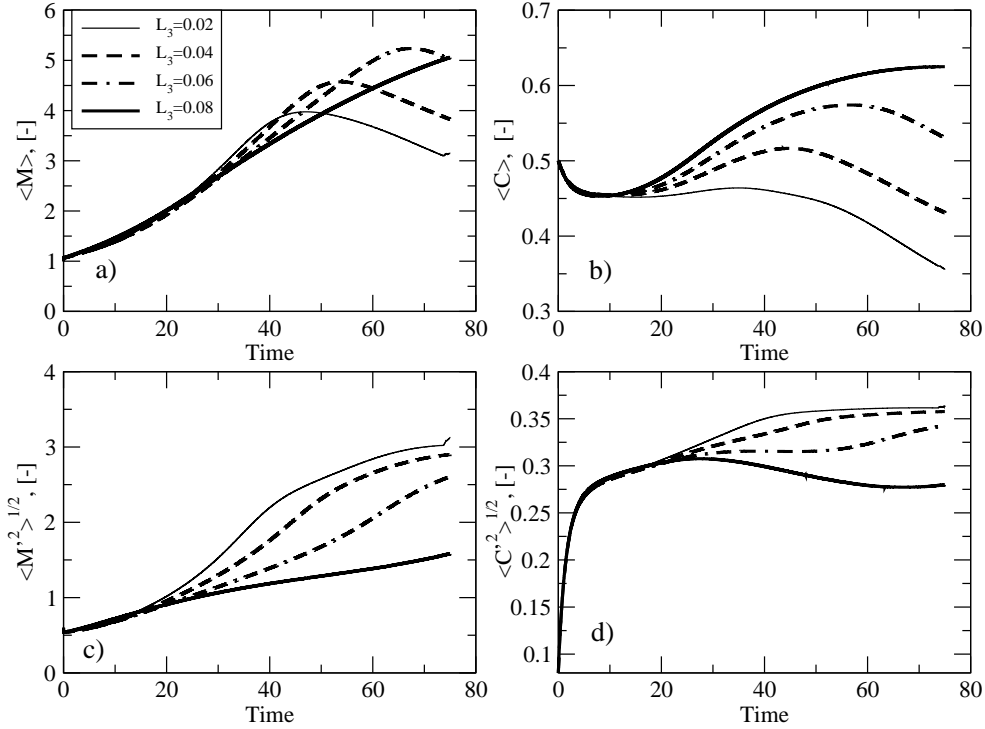}
\caption{Temporal evolution of the first two central moments of the wealth: a) average and c) root mean square,  and contentment: b) average and d) root mean square. The time, $t$, unit is one year. The values of the coefficient for the wealth tax in Eq.~\ref{TW} are shown in the legend.}%
\label{means}
\end{center}
\end{figure}
%\end{page}

\begin{figure}
\begin{center}
  \includegraphics[width=.48\textwidth]{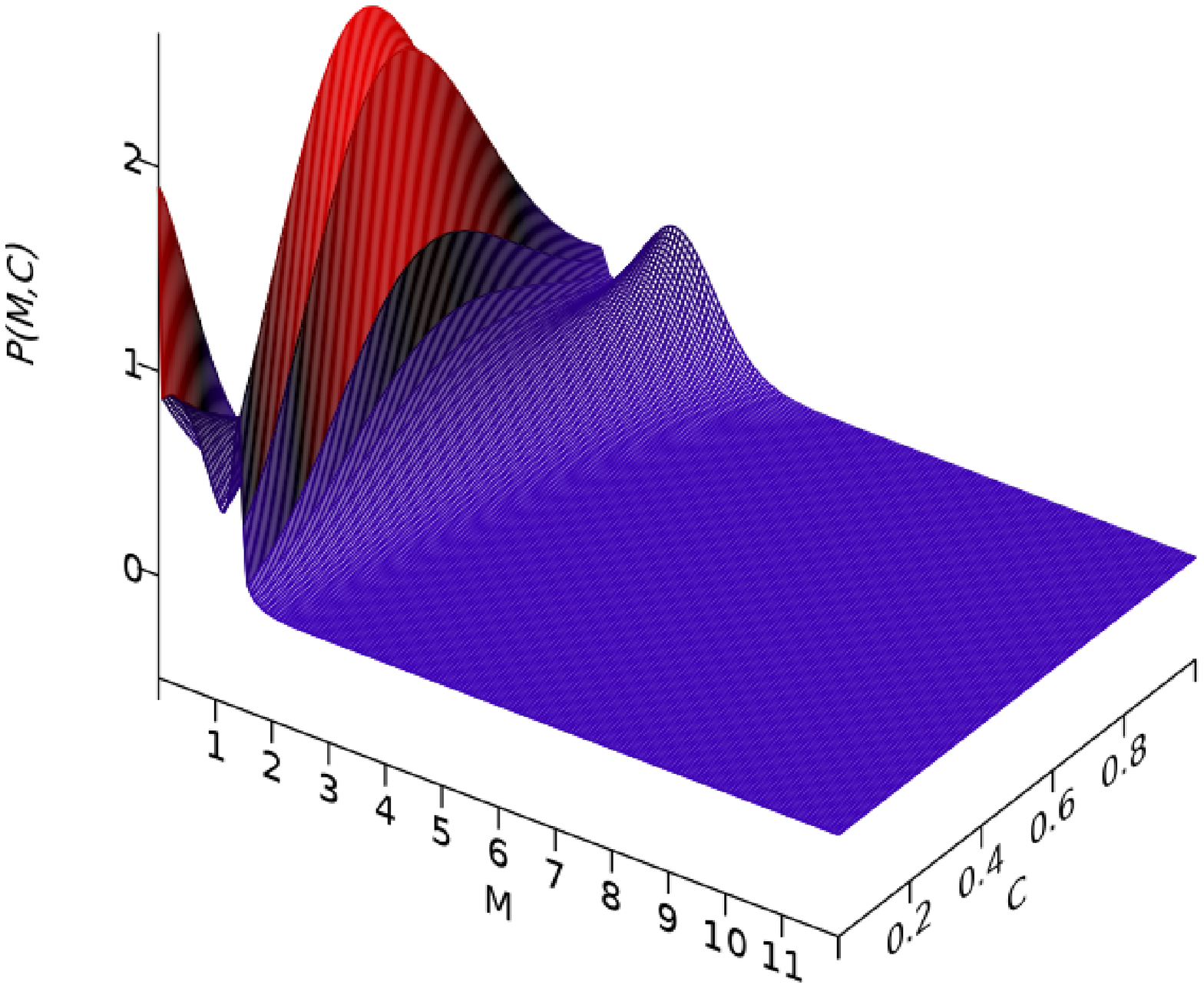}\hskip1.em
  \includegraphics[width=.48\textwidth]{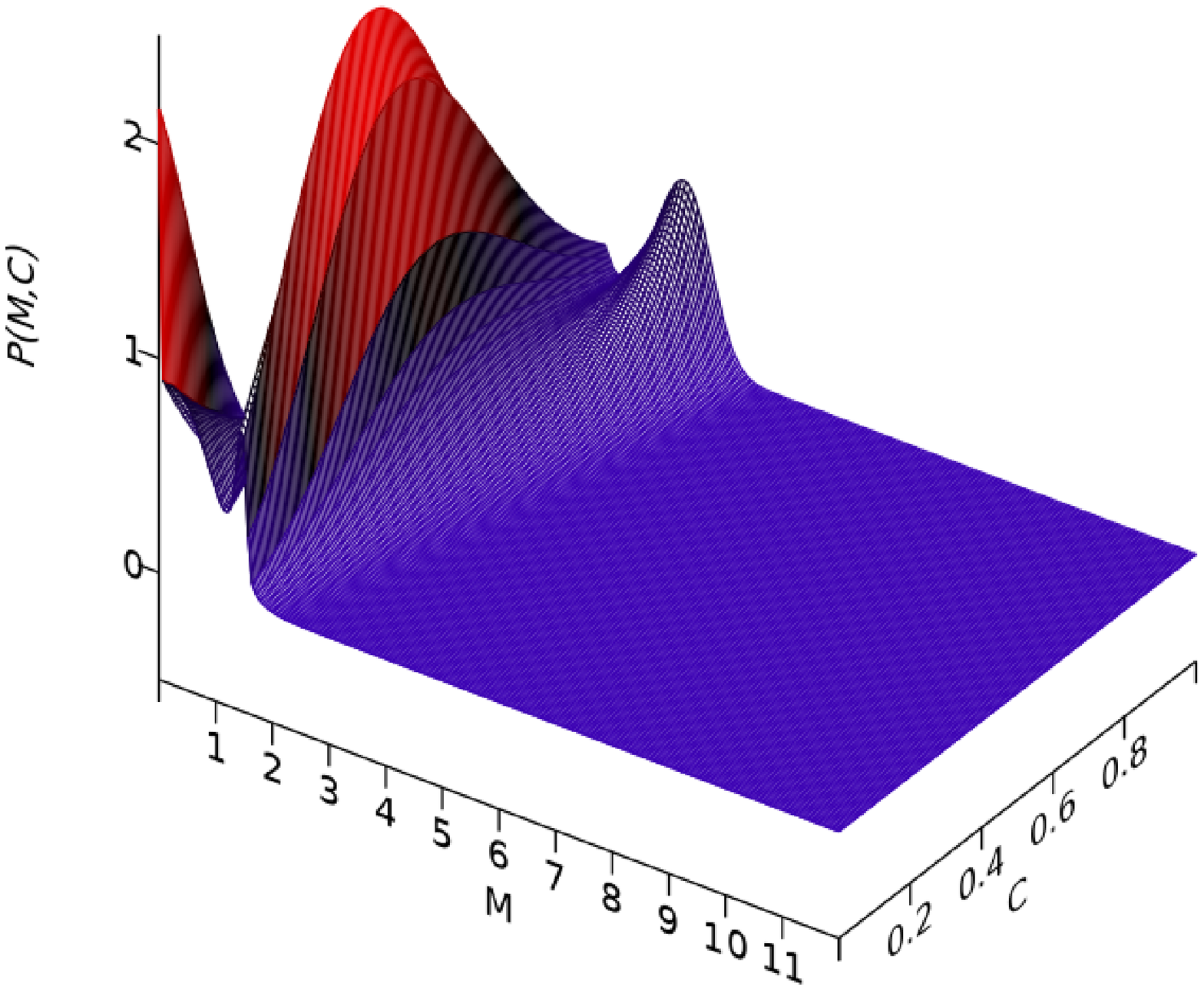}\hskip1.em
\caption{The joint pdf $P(M,C)$ of the wealth $M$ and contentment $C$  at time $t\approx 12$ for: left) low wealth tax, $L_3=0.02$, and right) high wealth tax, $L_3=0.08$.}%
\label{P11}
\end{center}
\end{figure}

\begin{figure}
\begin{center}
  \includegraphics[width=.48\textwidth]{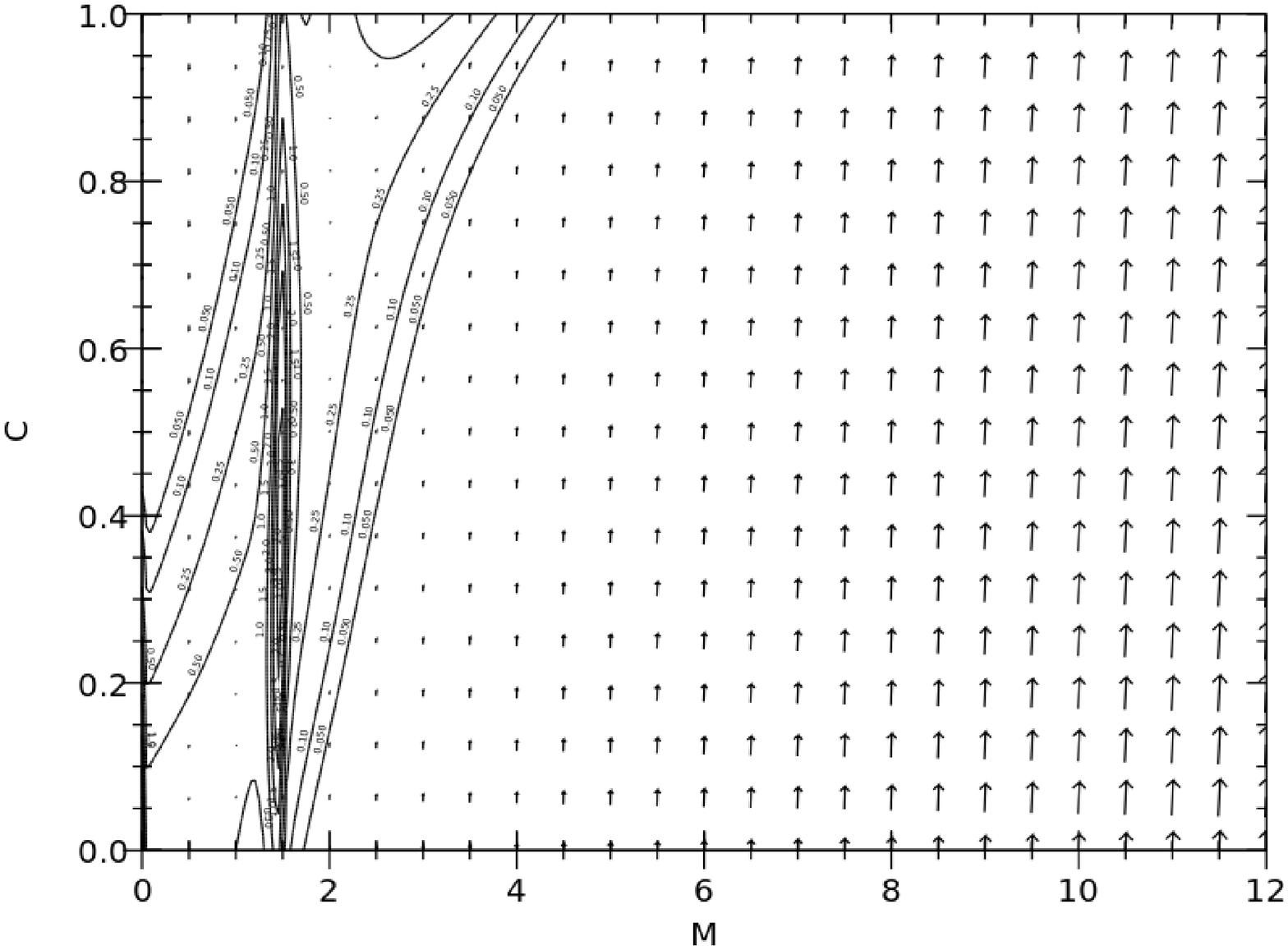}\hskip1.em
  \includegraphics[width=.48\textwidth]{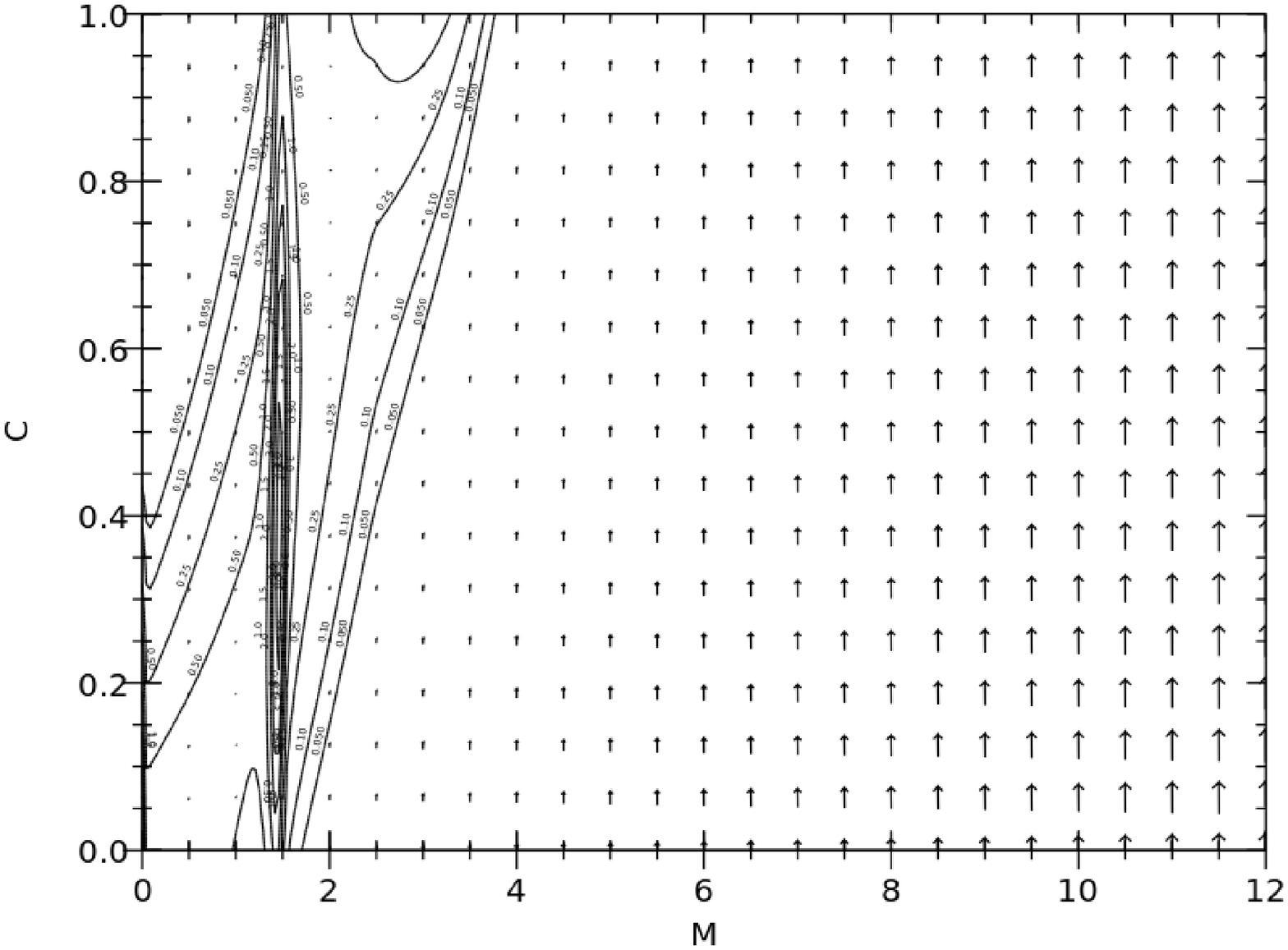}\hskip1.em
\caption{The flow in the wealth-contentment space induced by Eqs.~\ref{dCdt} and \ref{Mdot}  at time $t\approx 12$ for: left) low wealth tax, $L_3=0.02$, and right) high wealth tax at $L_3=0.08$. Values of $P(M,C)$, see Fig.~\ref{P11}, are shown on the superimposed isolines.}
\label{F11}
\end{center}
\end{figure}

\begin{figure}
\begin{center}
  \includegraphics[width=.46\textwidth]{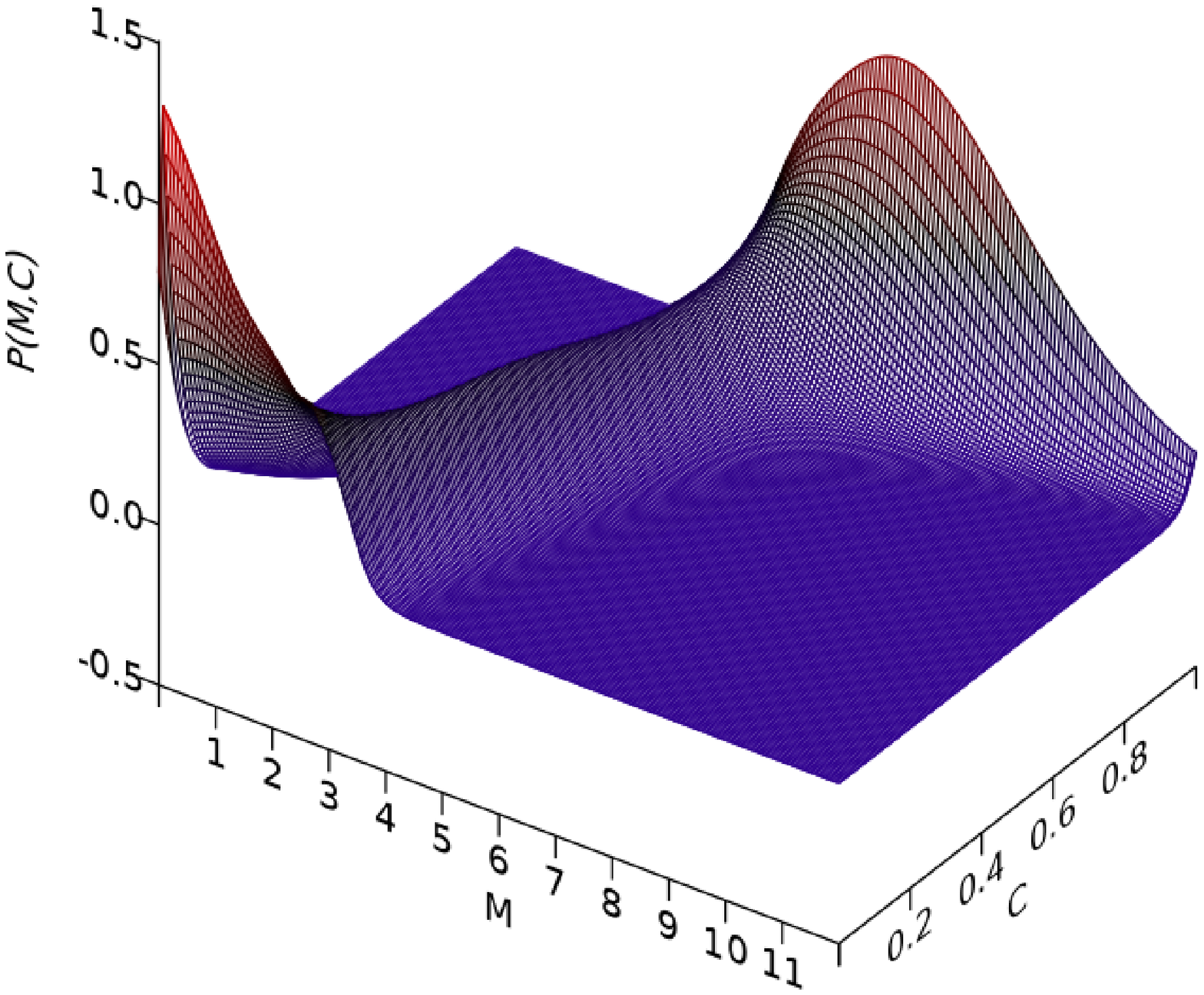}\hskip1.em
  \includegraphics[width=.50\textwidth]{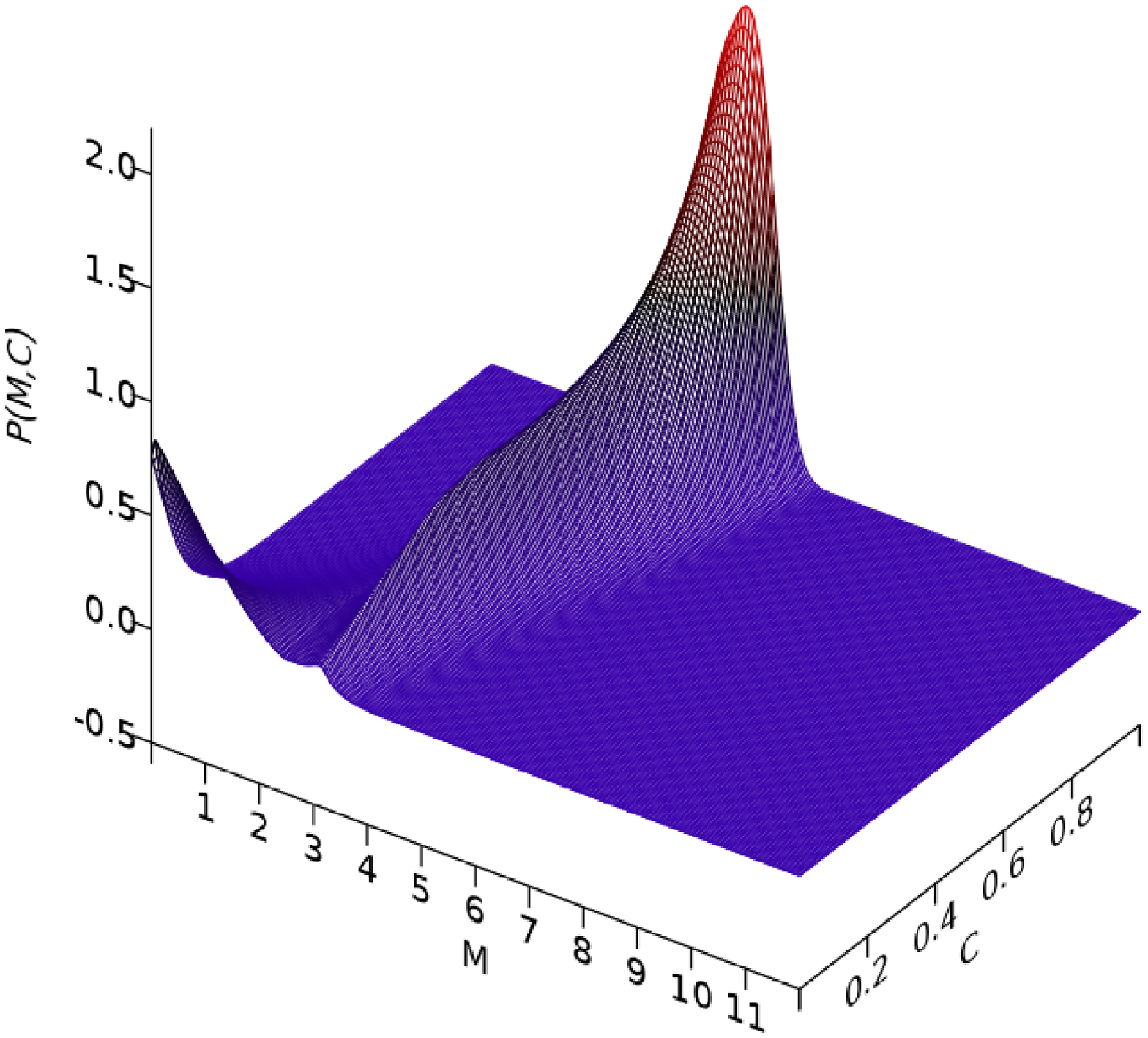}\hskip1.em
\caption{The joint pdf $P(M,C)$ of the wealth $M$ and contentment $C$ for: left) low wealth tax at $L_3=0.02$, time  $t\approx 38.3$ and right) high wealth tax at $L_3=0.08$, time $t\approx 38.3$.}%
\label{P38}
\end{center}
\end{figure}

\begin{figure}
\begin{center}
  \includegraphics[width=.48\textwidth]{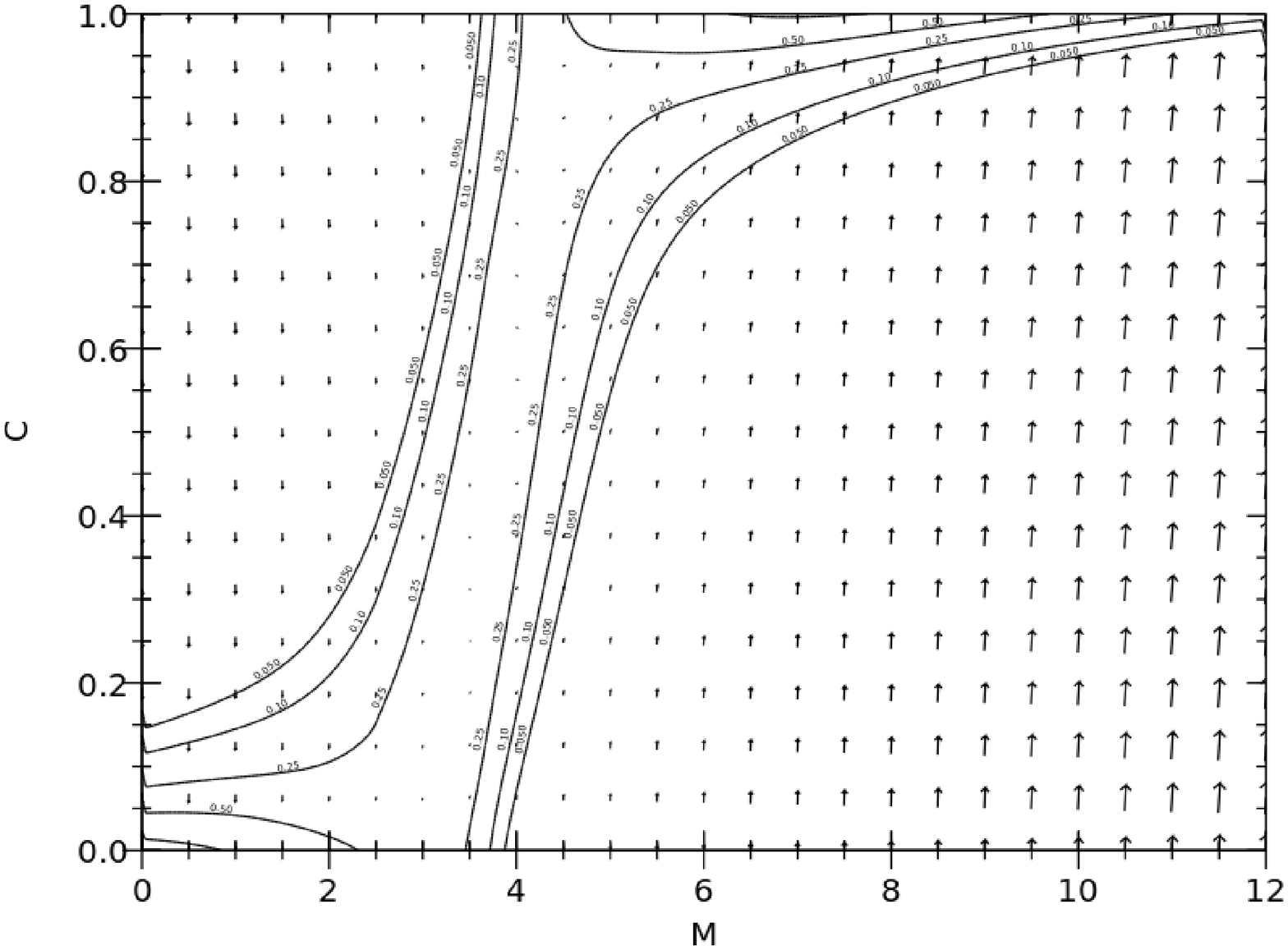}\hskip1.em
  \includegraphics[width=.48\textwidth]{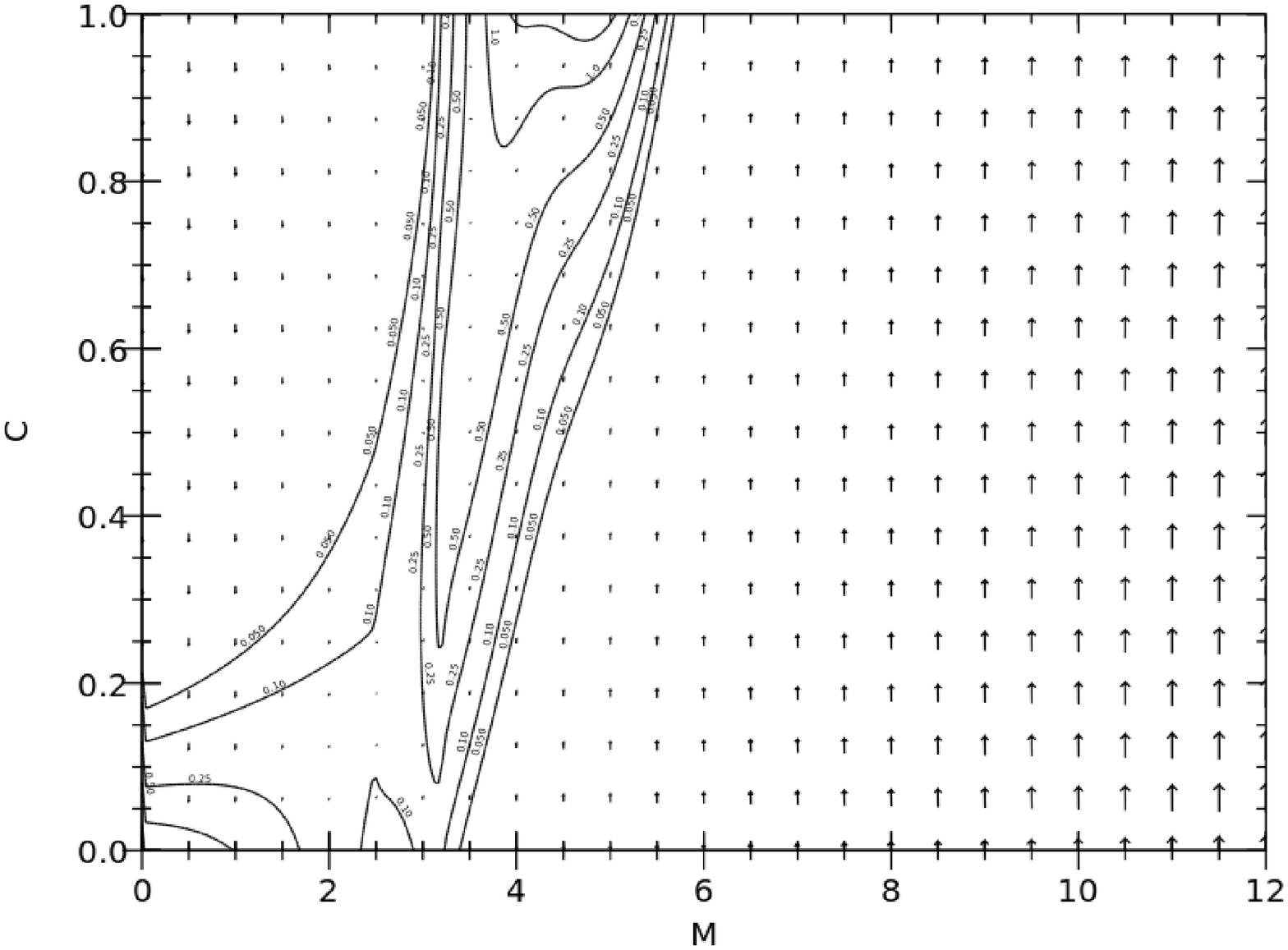}\hskip1.em
\caption{The flow in the wealth-contentment space induced by Eqs.~\ref{dCdt} and \ref{Mdot} for: left) low wealth tax at $L_3=0.02,\ t\approx 38.3$ and right) high wealth tax at $L_3=0.08, \ t\approx 38.3$. Values of $P(M,C)$, see Fig.~\ref{P38}, are shown on the superimposed isolines.}
\label{F38}
\end{center}
\end{figure}

In order to demonstrate its performance, the developed model is applied to this nearly homogeneous society and a parametric study is conducted of effects on contentment of the tax on wealth given by Eq.~\ref{TW}, the magnitude of which is governed by the $L_3$ parameter. The lower limit at which the wealth tax is levied is kept fixed at $M_w=2.5$. Figure \ref{means} shows the temporal evolution of the average and root-mean-square wealth and contentment for a four-fold variation of the rate of the wealth tax. Initially, for the first approximately 25 years or one change of generations, the wealth tax magnitude has very little influence: the the chosen initial $P(M,C)$ corresponds to a very small fraction of individuals with $M \ge M_w$. This is corroborated with Figs. \ref{P11} and \ref{F11} showing the $P(M,C)$ and its rate of change in the midst of this period for the smallest and largest wealth tax. Figure \ref{F11} shows the iso-contours of the $P(M,C)$ superimposed on the field of the its rate of change. One can see that the society undergoes stratification with appearance of noticeable fractions of individuals who are nearly entire destitute, see the rather sharp peak at $M=0, C=0$ but also those with more than twice the average wealth. Rather unsurprisingly, the model yields the destitute extremely unhappy and the rich very content. What is less  trivial is that, though both the average wealth, and the wealth of the largest fraction of the population are steadily increasing, this increase is accompanied with decreasing average contentment and formation of very wide contentment distribution, see Fig. \ref{F11}.

The model postulates that contentment increases the productivity, Eq.~\ref{U1}, thus non-zero $\ave{C'^2}^{1/2}$ leads to a modest growth of $\ave{M'^2}^{1/2}$, see Fig. \ref{means}c). The initial growth of $\ave{C'^2}^{1/2}$ is very fast owing to the combined effects of marriages, Eq.~\ref{Ifam}, random events, Eq.~\ref{W8},  and dependency of contentment on income, Eq. ~\ref{W2} and wealth Eq.~\ref{W1}. At this moment, the effects of the wealth tax rate $L_3$ are small: larger $L_3$ lead to small decrease of fraction of individuals with $M\approx 3, C\approx 1$.  

These effects, however, become more and more pronounced in longer run. Even though the average individual wealth increases slightly faster for smaller wealth tax for the first 40 years, it attains a plateau and decreases afterwards, while growth of $\ave{M}$ with larger wealth tax is continuous for the entire period of simulations, cf. thin and thick solid lines in Fig.~\ref{means}a). Rather curiously, lower wealth tax with its faster growth of $\ave{M}$ does not yield proportional increase in average contentment: for larger $L_3$, after the initial decrease $\ave{C}$ stagnates and then decreases while it grows steadily for larger $L_3$, so that the final values of average contentment are about twice larger for larger wealth tax, cf. thin and thick solid lines in Fig.~\ref{means}b). Larger rate of the wealth tax also decreases stratification of the society meaning much slower growth and smaller values of both root-mean-square wealth and contentment, cf.  thin and thick solid lines in Fig.~\ref{means}c) and d). 

Figure \ref{P38} illustrates the difference in the stratification of the society for the instant when the smaller wealth tax leads to the stagnation in growth of the average wealth. One may clearly see that, while unsurprisingly there is larger probability of larger, $M> M_w$, individual wealth, there is also very much larger fraction of the society in the destitution and discontent $M\approx 0, C\approx 0$. Combination of two factors explain this increased stratification: the first relates to the appropriation of the added value by the rich, Eq.~\ref{U2}, and the second is that the redistributive action of marriage is ineffectual for reducing the wealth extremes when their difference exceeds two rms wealth values, cf. Eq.~\ref{fz}. Even though the society at this instant is about three times wealthier in average than it was at the initial time, the stratification caused by smaller wealth tax yields appreciably smaller average contentment. It may be inferred from Fig.~\ref{F38} that the added value is alienated by a progressively smaller from a progressively larger fraction of the society. 

\begin{figure}
\begin{center}
  \includegraphics[width=.48\textwidth]{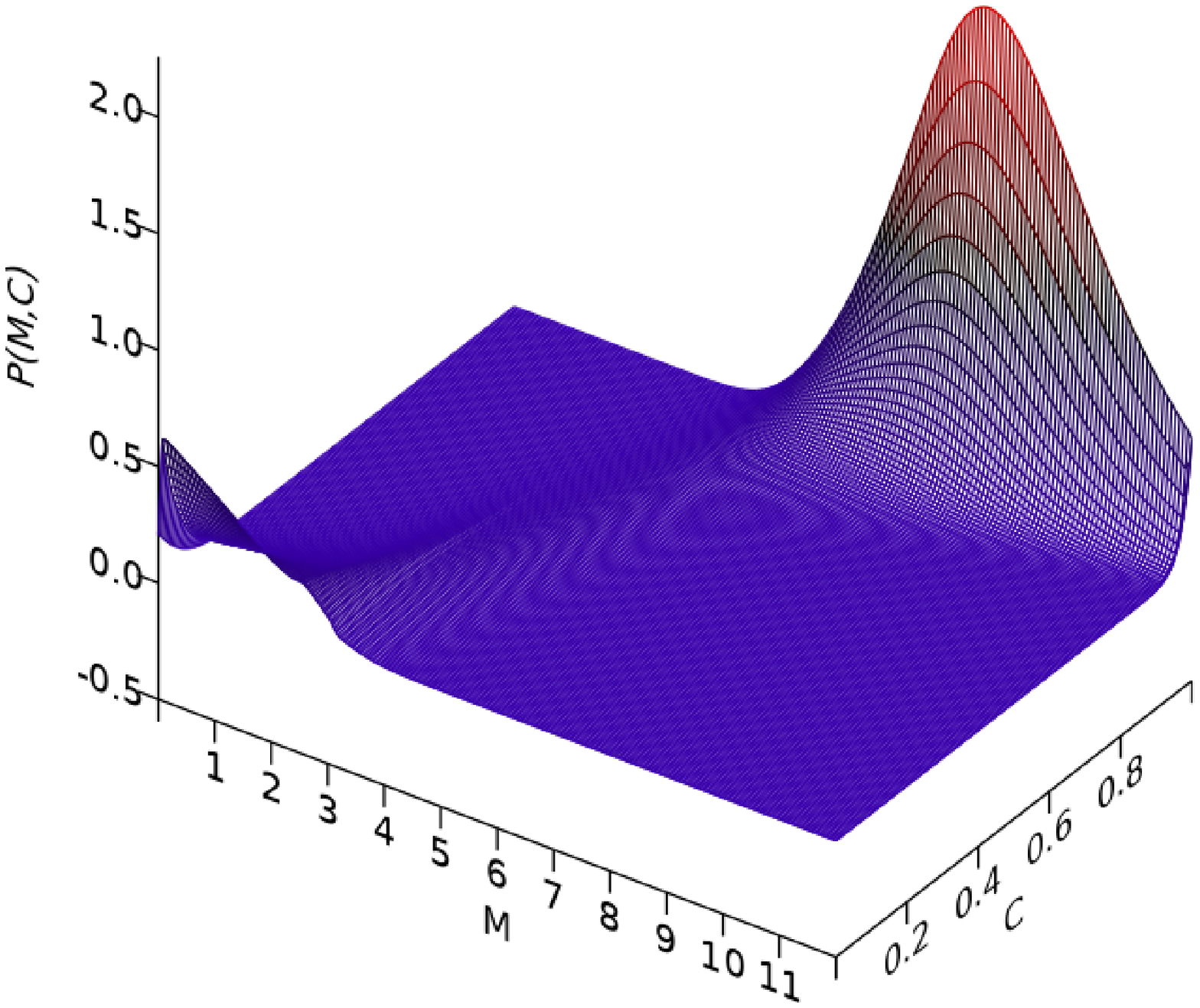}\hskip1.em
  \includegraphics[width=.47\textwidth]{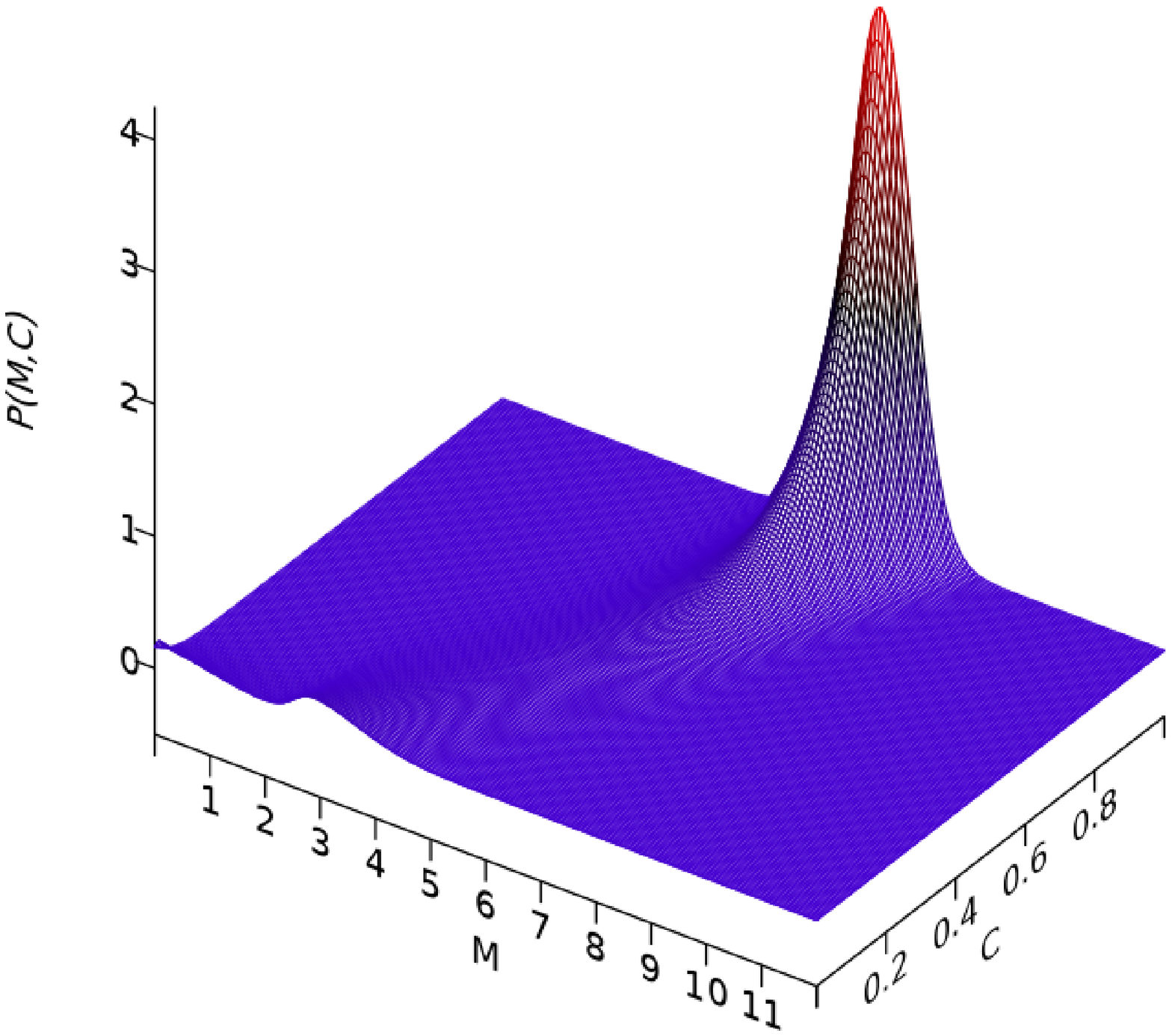}\hskip1.em
\caption{The joint pdf $P(M,C)$ of the wealth $M$ and contentment $C$ for: left) low wealth tax at $L_3=0.02,\ t\approx 72.4$ and right) high wealth tax at $L_3=0.08, \ t\approx 72.9$.}%
\label{P72}
\end{center}
\end{figure}

\begin{figure}
\begin{center}
  \includegraphics[width=.48\textwidth]{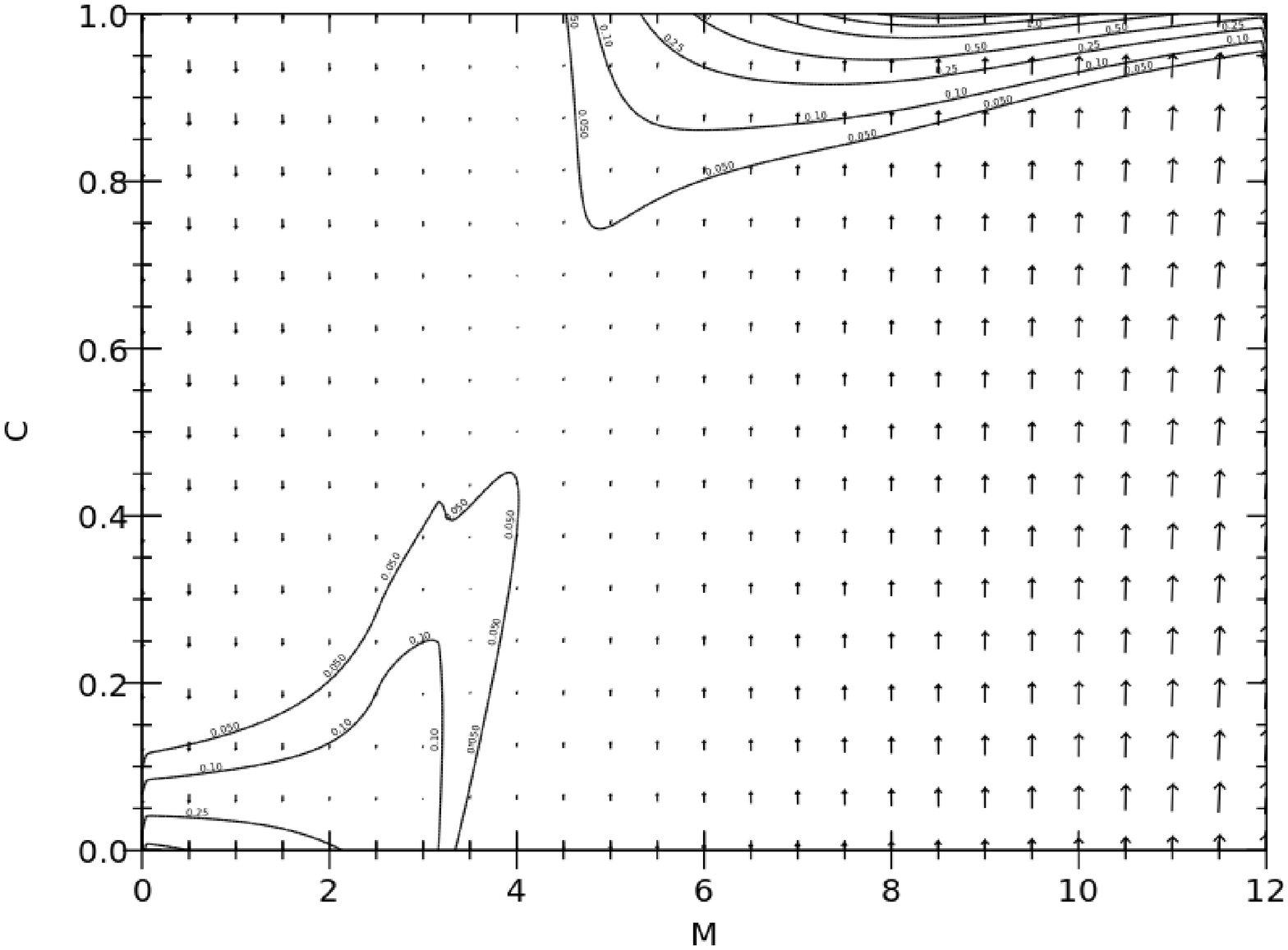}\hskip1.em
  \includegraphics[width=.48\textwidth]{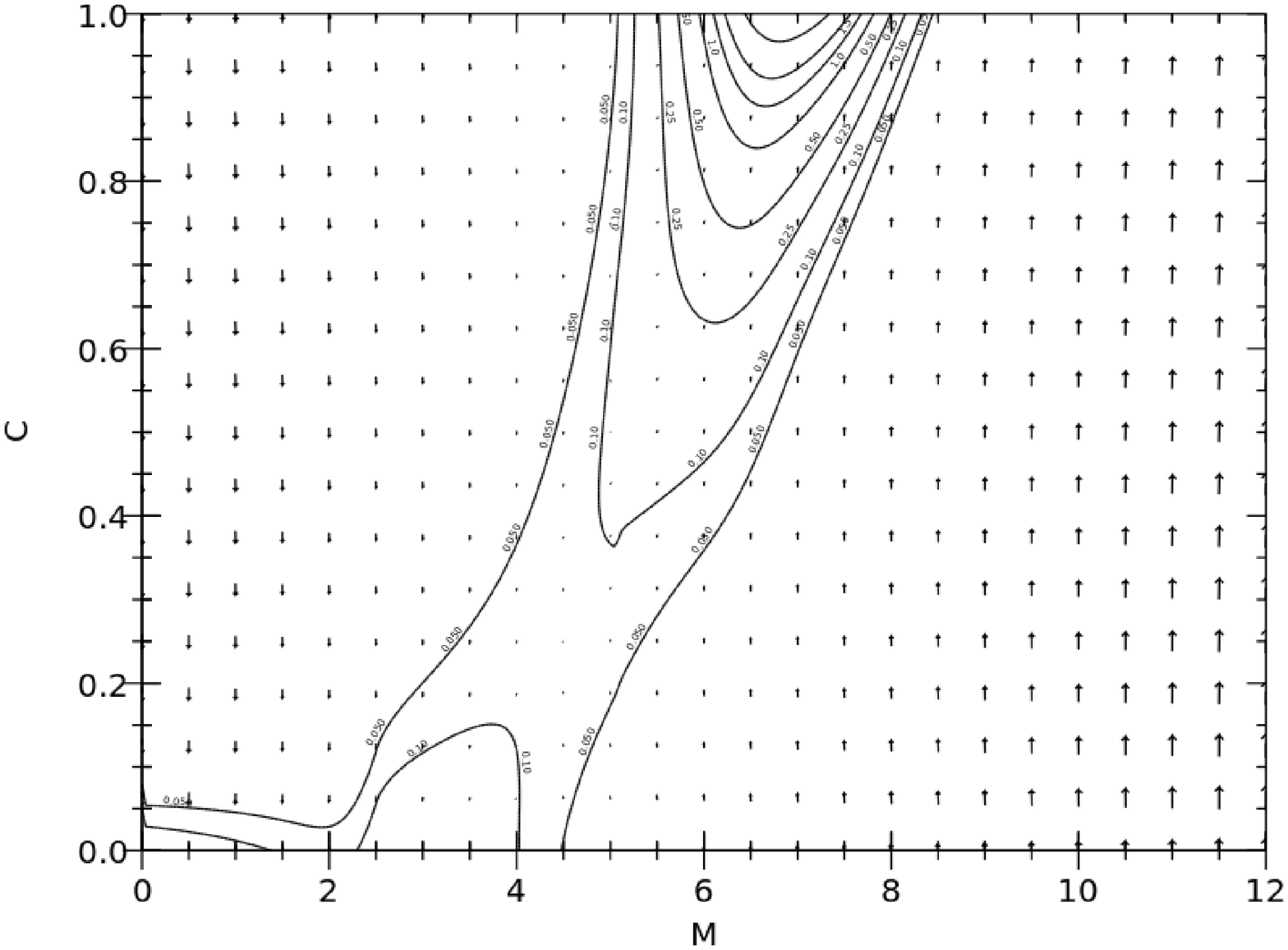}\hskip1.em
\caption{The flow in the wealth-contentment space induced by Eqs.~\ref{dCdt} and \ref{Mdot} for: left) low wealth tax at $L_3=0.02,\ t\approx 72.6$ and right) high wealth tax at $L_3=0.08, \ t\approx 72.9$. Values of $P(M,C)$, see Fig.~\ref{P38}, are shown on the superimposed isolines.}
\label{F72}
\end{center}
\end{figure}

These trends continue and towards the end of simulation, $t\approx 75$, the low wealth tax case show a fully stratified society, see Figs.~\ref{P72} and  ~\ref{F72}. The high wealth tax society is more homogeneous, with zero probability of very high individual wealth $M\ge 9$ and much smaller fraction of the destitute; Fig.~\ref{means} shows that this case has much higher average wealth and contentment. Comparison of the temporal evolution of $P(M,C)$ clearly shows that there is no simple relationship between the wealth and contentment and only the average values of these variables are not sufficient to predict further evolution of the society. This provides a quantitative explanation to the Easterlin paradox \citep{East10}. An interesting feature of the $P(M,C)$ distribution in this case is the fairly wide spread of contentment among the well-off part of the population with  noticeable probability of meeting simultaneously wealth well above and contentment well below the average values for the society as a whole. 
%\subsection*{ $P(M,C)$}

\section*{Conclusions}

Use of a variable continuously changing between 0 and 1 for contentment, or satisfaction with life, of an individual allows to formulate an ordinary differential equation governing its temporal evolution. This equation includes 
several objective and subjective factors of which the influence is expressed through simple expressions and estimations of their characteristic times. 
From this individual's contentment equation, and coupled to it  equation for
temporal evolution of the individual's wealth, it proves possible to formulate a mathematical model, Eq.~\ref{P2}, describing temporal evolution of the joint probability density of individuals' wealth and contentment within a society.

The model is constructed with resort to fairly general arguments about time-scales of influence of different factors on contentment and with invocation of very simple expressions for action of those factors. Arguably, the individual's wealth equation is better founded as its terms reflect the current productivity and tax regimes in the UK, still the representation is simplified. Despite this very approximate reasoning, the model yields results  which are at least qualitatively plausible. It should be stressed that alternative expressions for any of the factors affecting contentment (or wealth) may be easily incorporated into the model without affecting its underlying principles. It is hoped that the proposed model could be of use in further quantitative works clarifying statistical characteristics of the satisfaction with life. 

The first application of the evolution equation for the joint distribution of wealth and contentment is made here to illustrate the effects of tax on the wealth on the society. It is shown that the magnitude of this particular form of tax has a significant impact on the degree of stratification in the society, in terms of both contentment and wealth, and the less stratified society is, in long run, wealthier and its members have generally the greater contentment.

\section*{Acknowledgements}
The impetus to this work was given by discussions with Prof. Roland Borghi whose influence on the author cannot be overestimated.

\bibliographystyle{model1-num-names}
\bibliography{content}

\end{document}